\documentclass[pdftex]{webofc}
\usepackage[varg]{txfonts}
\RequirePackage[colorlinks,citecolor=blue,urlcolor=blue,linkcolor=blue]{hyperref}
\RequirePackage{amsfonts}
\RequirePackage{amsmath,amssymb}
\RequirePackage{bm}
\usepackage{fancyhdr, graphicx}
\usepackage{float}
\usepackage{color}
\usepackage{amsmath,amssymb,wasysym}
\usepackage{multirow}
\usepackage{epstopdf}
\usepackage{lineno}

\setbox123\hbox{\small$0$}
\def\S{\hbox to\wd123{\hss}}
\setbox124\hbox{\small$_{0}$}
\def\s{\hbox to\wd124{\hss}}

\begin{document}

\title{Convergence properties of L\'evy expansions: 
implications for Odderon and proton structure}

\author{
       \firstname{T.} \lastname{Cs\"{o}rg\H{o}}\inst{1,2,3}\fnsep\thanks{\email{tcsorgo@cern.ch}} 
       \firstname{R.} \lastname{Pasechnik}\inst{4}\fnsep\thanks{\email{Roman.Pasechnik@thep.lu.se}}
       \and
       \firstname{A.} \lastname{Ster}\inst{1}\fnsep\thanks{\email{Ster.Andras@wigner.mta.hu}}
}

\institute{MTA Wigner FK, H-1525 Budapest 114, P.O.Box 49, Hungary
          \and
          EKE KRC, H-3200 Gy\"ongy\"os, M\'atrai \'ut 35, Hungary 
          \and
          CERN, CH - 1211 Geneva 23, Switzerland
          \and
          Department of Astronomy and Theoretical Physics, 
          S\"olvegatan 14 A,  SE - 223 62 Lund, Sweden
          }
  
\abstract{
We detail here the convergence properties of a new model-independent
imaging method, the  L\'evy expansion, that seems to play an important role in the analysis of the differential 
cross section of elastic hadron-hadron scattering. We demonstrate, how our earlier results concerning 
the Odderon effects in the differential cross-section of elastic proton-proton and proton-antiproton
scattering as well as those related to apparent sub-structures inside the protons were obtained in a convergent and stable manner.
}

\maketitle

\section{Introduction}
\label{s:intro}

The model-independent L\'evy imaging technique ~\cite{Csorgo:2018uyp,Csorgo:2018ruk,Csorgo:2019rsr}
has recently become a useful, simple and unambiguous tool for extracting the physics information from the elastic 
hadron-hadron scattering data in a statistically acceptable manner. The power of this technique has been first 
demonstrated in our earlier analysis of the most recent data sets from the total, elastic and differential 
cross-section measurements in elastic $pp$ collisions $\sqrt{s} = 13$ TeV performed by the TOTEM Collaboration 
at the Large Hadron Collider (LHC) (for a few most recent TOTEM publications, see Refs.~\cite{Antchev:2017dia,Antchev:2017yns,Antchev:2018edk,Antchev:2018rec}). These TOTEM results,
and in particular the  comparison of the differential cross-section of elastic proton-proton scattering
at $\sqrt{s} = 2.76$ TeV with D0 results on elastic proton-antiproton scattering at 1.96 TeV~\cite{Abazov:2012qb}
indicate several Odderon effects, as discussed recently in Refs.~\cite{Csorgo:2018uyp,Antchev:2018rec}

In particular, indirect signatures of the Odderon exchange in differential elastic $pp$ and $p\bar p$ cross-sections 
have been identified by using the L\'evy imaging technique, also known as the model-independent L\'evy expansion method. 
Another important implication of this technique is that it enables to probe the internal structure of the proton by identifying 
its smaller substructures imprinted in the behaviour of the $t$-dependent elastic slope $B(t)$. In particular, the proton 
substructures of two distinct sizes in the low (a few tens of GeV) and high (a few TeV) energy regimes, respectively, have been 
found and discussed in Refs.~\citep{Csorgo:2018uyp,Csorgo:2018ruk,Csorgo:2019rsr}. 
A remarkable feature of the L\'evy expansion of the elastic amplitude is that the diffractive
cone is described fairly well by the L\'evy-stable (or stretched exponential) distribution in terms of two free parameters only, 
the L\'evy scale parameter $R$ characterising the length-scale of the scattered systems, and the exponent $\alpha=\alpha_L/2$. One
of the conventions of elastic scattering is to parameterize the diffractive cone with an exponential behaviour,
corresponding to $\alpha = 1$, i.e. to a Gaussian scattering amplitude in the impact parameter space.
The exponent $\alpha_L = 2 \alpha $ is known as the L\'evy index of stability providing a small 
but significant deviation of the cone shape from the classically expected Gaussian shape. A detailed presentation of this 
sophisticated new imaging technique is detailed in Ref.~\cite{Csorgo:2018uyp}, while for a brief summary of the main results, 
see Refs.~\cite{Csorgo:2018ruk,Csorgo:2019rsr}.

Given a power of the L\'evy imaging technique operating with very few initial assumption for description of a large amount of data, 
a natural question arises about the stability and convergence of the associated L\'evy series that are used for mapping the elastic 
amplitude with this method. In this contribution, we attempt to give a short description of the main results of the method, as well as 
demonstrate its convergence properties for a large variety of data sets at different scattering energies. The main results were
summarized recently in two short contributions~\cite{Csorgo:2018ruk,Csorgo:2019rsr}, however the convergence properties
of this model-independent L\'evy expansion method were not yet detailed in the literature before.

\section{Model-independent L\'evy expansion of the scattering amplitude}
\label{s:Levy-expansion}

\begin{figure*}[!h]
 \begin{center}
 {\includegraphics[width=1.0\textwidth]{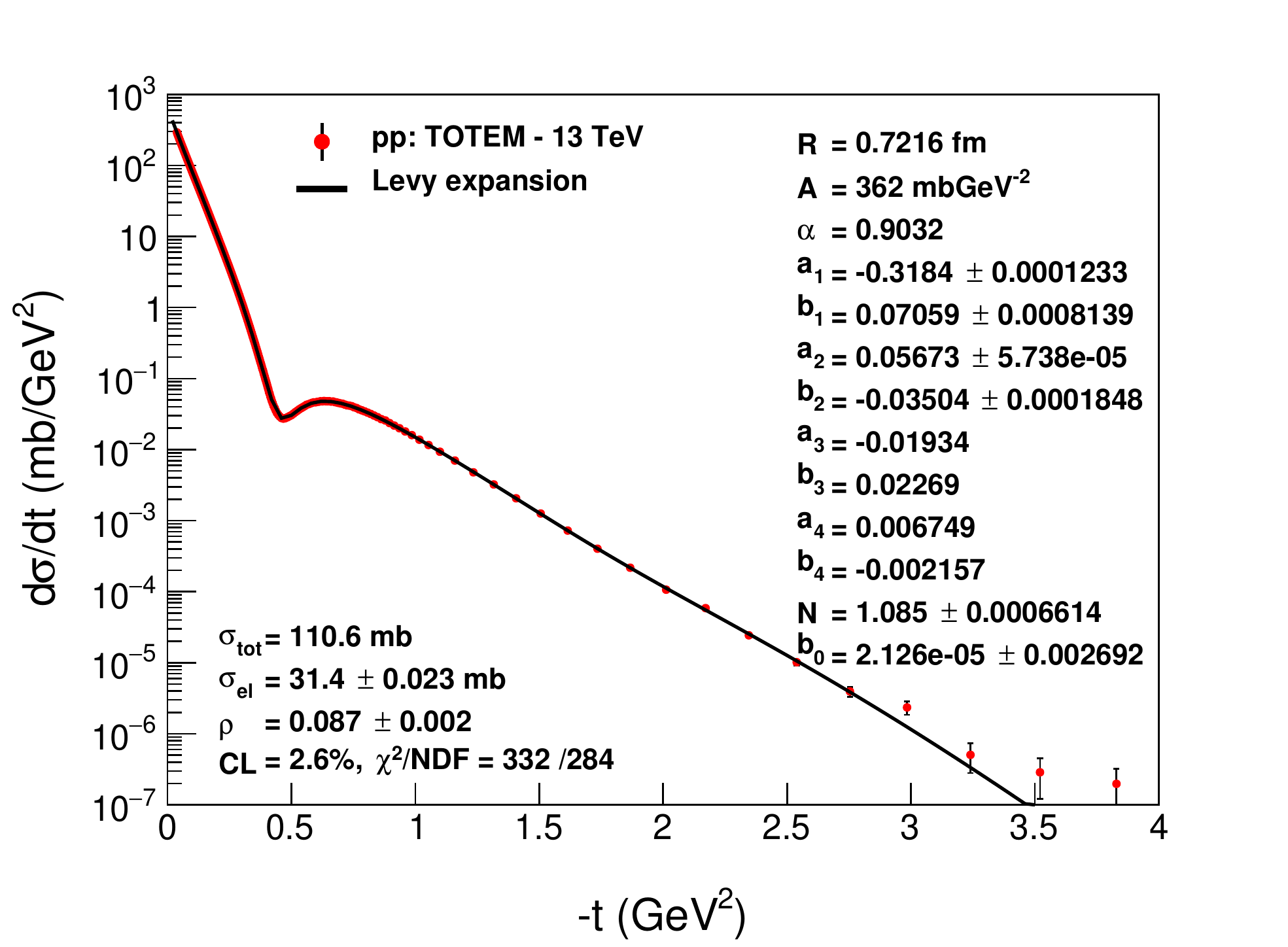}}
 \end{center}
\caption{
Fourth order L\'evy expansion to $\sqrt{s} = 13 $ TeV $pp$ elastic scattering data
using Eq. ~(\ref{e:dsigmadt-Tel}) with a $b_0$ as a free parameter. The comparision to the TOTEM data
from Ref.~\cite{Antchev:2018edk} resulted in a statistically acceptable description within quadratically
added statistical and systematic errors. The fit parameter $b_0$ was within errors found to be zero, hence
it was fixed to zero in the subsequent analysis.
}
\label{f:b0}
\end{figure*}

In order to study systematically the deviations of a given data set on the elastic cross section $d\sigma/dt$, differential in four-momentum 
transfer $t=(p_1 - p_3)^2<0$, from an approximate L\'evy-stable shape apparent at larger $t$ beyond the diffractive cone, we adopt the
L\'evy series expansion for the elastic amplitude $T_{\rm el}(\Delta)$, $\Delta=\sqrt{|t|}$, represented in terms of a complete orthonormal set of 
L\'evy polynomials (with complex coefficients) and a L\'evy weight function $w(z|\alpha) =  \exp(-z^\alpha)$~\cite{Csorgo:2018uyp,Csorgo:2018ruk}. 
While the L\'evy polynomials exhibit an oscillatory behaviour in dimensionless scaling variable $z=|t| R^2$, the differential cross-section of elastic 
scattering is proportional to a hit distribution which is, by construction, a positively-definite function of $t$, i.e.
\begin{eqnarray}
\frac{d\sigma}{dt} = \frac{1}{4\pi} \, |T_{\rm el}(\Delta)|^2 \,, \quad
T_{\rm el}(\Delta) &=& i\sqrt{4\, \pi \, A\,  w(z|\alpha)}\, 
		\left[1 \, + i b_0 \, + \, \sum_{i = 1}^\infty c_i l_i (z|\alpha) \right] \, , 
\label{e:dsigmadt-Tel}
\end{eqnarray}
where $c_j=a_j + i b_j$ are the complex coefficients of the L\'evy expansion ($j = 0, 1, ... , )$ and $a_0 = 1$ fixed so that the overall normalization can be absorbed into the coefficient $A$. This simple expansion form indicates three underlying physical
assumptions, with further details explained below.
\begin{enumerate}
    \item The leading order behaviour corresponds to a nearly exponential (in other words, a non-exponential) distribution. This leading order 
    behaviour is $\frac{d\sigma}{dt} = A \exp\left[ - (R^2|t|)^\alpha\right]$, which is consistent with the TOTEM observation of
    a non-exponential low-$|t|$ behaviour of the differential cross-section of elastic $pp$ scattering at $\sqrt{s} = 8$ TeV ~\cite{Antchev:2015zza}. 
    In the limit of $\alpha$ $\rightarrow$ 1, the exponential cone behaviour
    is recovered, with $\lim_{\alpha\rightarrow 1} \frac{d\sigma}{dt} = A \exp\left[ - B|t|\right]$ with $B = R^2$.
    \item The leading order scattering amplitude is assumed to be imaginary, with vanishing real part, corresponding to $b_0 = 0$ and
    to all the $c_j$ expansion coefficients vanish for $j \ge 1$.  Such an assumption can be relaxed by assuming that $b_0$ can be different from zero, but so far all the fits that we have performed, we have found 
    $b_0$ to be consistent within errors with zero, so we have fixed this possible expansion parameter to zero, accordingly. 
    The possibility of relaxing $b_0$ to be different from zero
    is investigated in Fig.~\ref{f:b0} and it is briefly discussed there in the context of the $\sqrt{s} = 13$ TeV elastic $pp$ scattering data.
    \item For predominantly imaginary scattering amplitudes with $b_0 = 0$, the existence of a non-vanishing real to imaginary ratio $\rho(t)$,
    in particular, $\rho_0 = \rho(t=0)$, is mapped one-to-one to the existence of a diffractive interference structure, i.e.
    a diffractive minimum in elastic scattering cross section. Given that for $t \ne 0$ the L\'evy expansion is an analytic function, and that at the first
    diffractive minimum the imaginary part is expected to vanish, the phase of the scattering amplitude can be uniquely determined with the
    L\'evy expansion method up to an overall sign, that can be fixed from measurements at the Coulomb-nuclear interference region.
    This implies that the $\rho_0$ parameter can be determined from the L\'evy expansions if the fit range includes a diffractive minimum.
\end{enumerate}

The set of orthonormal polynomials are denoted above by $\left\{l_j(z|\alpha)\right\}_{j= 0}^\infty$, while
the set of orthogonal, but unnormalized L\'evy polynomials are denoted by $\left\{L_j(z|\alpha)\right\}_{j= 0}^\infty$.
These unnormalized L\'evy polynomials  were introduced in Refs.~\cite{DeKock:2012gp,Novak:2016cyc}, to analyze nearly L\'evy 
shaped Bose-Einstein correlations in two-particle Bose-Einstein correlations or particle interferometry. 
Their orthogonality is defined with respect to  the weight function $w(z|\alpha)= \exp\left( - |z|^\alpha\right)$ 
where the dimensionless variable $z = \vert t\vert R^2$. The weight function acts simultaneously as the leading order approximate 
form of the measured distribution, and at the same time also as a measure in an abstract Hilbert
space, where the convergence properties of the L\'evy series can be investigated, following the general ideas proposed in
Ref.~\cite{Csorgo:1999wx}. Thus, this expansion method allows to keep the number of expansion coefficients at the minimal necessary level.

The orthonormality of the  L\'evy polynomials $l_n(z\,  |\, \alpha)$  is expressed as
\begin{eqnarray}
    \int_0^{\infty} dz \exp(-z^\alpha) \, l_n(z\,  |\, \alpha) \, l_m(z\,  |\, \alpha) & = & \delta_{n,m} \,.
    \label{e:lj}
\end{eqnarray}
These orthonormalized polynomials $l_j(z\, |\alpha)$ are proportional to 
the orthogonal -- but unnormalized ---  L\'evy polynomials, $L_j(z\, |\, \alpha)$,
\begin{eqnarray} 
  l_{j}(z\, |\, \alpha) & = &  D^{-\frac{1}{2}}_{j}(\alpha) \, D^{-\frac{1}{2}}_{j+1}(\alpha) \, 
    	L_j(z\, |\, \alpha) \, , \qquad \mbox{\rm for } j \ge 0,
\end{eqnarray}
which in turn are  given by a Gram-Schmidt orthogonalization procedure as
\begin{eqnarray}
 L_0(z\,|\,\alpha) & =&  1 , \\
 L_1(z\,|\,\alpha)  &= & 
       \det\left(\begin{array}{c@{\hspace*{8pt}}c}
     \mu_{0,\alpha} & \mu_{1,\alpha}  \\ 
     1 & z \end{array} \right) , \\
 L_2(z\,|\,\alpha) & =& 
       \det\left(\begin{array}{c@{\hspace*{8pt}}c@{\hspace*{8pt}}c}
     \mu_{0,\alpha} & \mu_{1,\alpha} & \mu_{2,\alpha} \\ 
     \mu_{1,\alpha} & \mu_{2,\alpha} & \mu_{3,\alpha}  \\ 
     1 & z & z^2 \end{array} \right), \quad \dots \; \mbox{\rm etc} \,.
\end{eqnarray}
introduced previously in Ref.~\cite{DeKock:2012gp}. The Gram-determinants of order $j$, 
$D_j \equiv D_j(\alpha)$ defined as
\begin{eqnarray}
D_0(\alpha) & =&  1 , \\
D_1(\alpha)  & =&   \mu_{0,\alpha} , \\
D_2(\alpha)  & = &
       \det\left(\begin{array}{c@{\hspace*{8pt}}c}
     \mu_{0,\alpha} & \mu_{1,\alpha}  \\ 
     \mu_{1,\alpha} & \mu_{2,\alpha} 
       \end{array} \right) , \\
D_3(\alpha) & =& 
       \det\left(\begin{array}{c@{\hspace*{8pt}}c@{\hspace*{8pt}}c}
     \mu_{0,\alpha} & \mu_{1,\alpha} & \mu_{2,\alpha} \\ 
     \mu_{1,\alpha} & \mu_{2,\alpha} & \mu_{3,\alpha} \\ 
     \mu_{2,\alpha} & \mu_{3,\alpha} & \mu_{4,\alpha} 
       \end{array} \right), \quad \dots \; \mbox{\rm etc.} 
       \label{e:gram3}
\end{eqnarray}
In the above expressions, 
\begin{eqnarray}
\mu_{n,\alpha} = \int_0^\infty dz\;z^{n} \exp( - z^\alpha) = \frac{1}{\alpha}\,\Gamma\left( \frac{n+1}{\alpha}\right) \,, \quad
\Gamma(x) = \int_0^\infty dz\;z^{x-1}e^{-z} \, ,
\end{eqnarray}
where $\Gamma(x)$ the Euler's gamma function.
\begin{figure*}[!h]
\begin{minipage}{0.495\textwidth}
 \centerline{\includegraphics[width=1.0\textwidth]{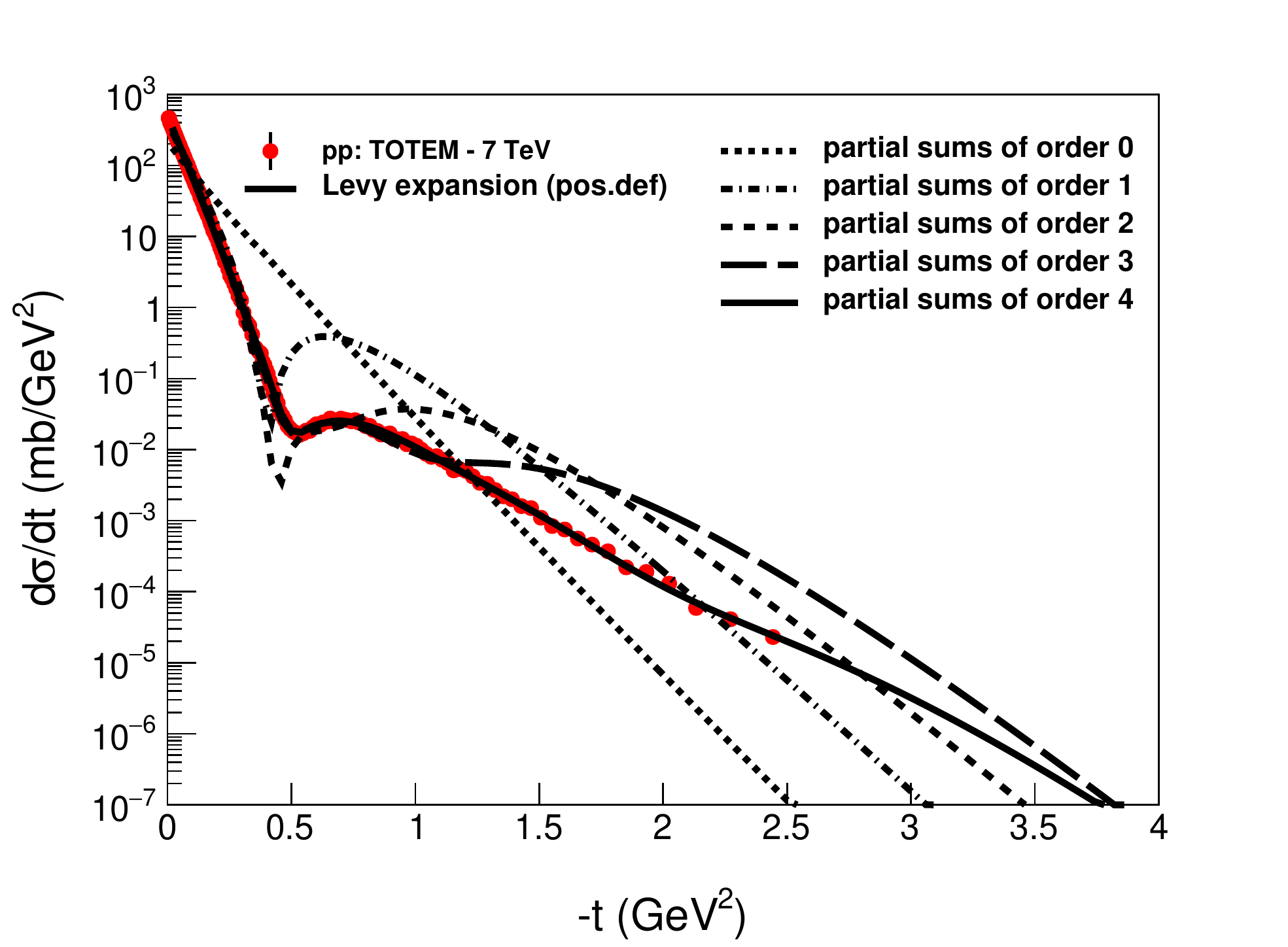}}
\end{minipage}
\begin{minipage}{0.495\textwidth}
 \centerline{\includegraphics[width=1.0\textwidth]{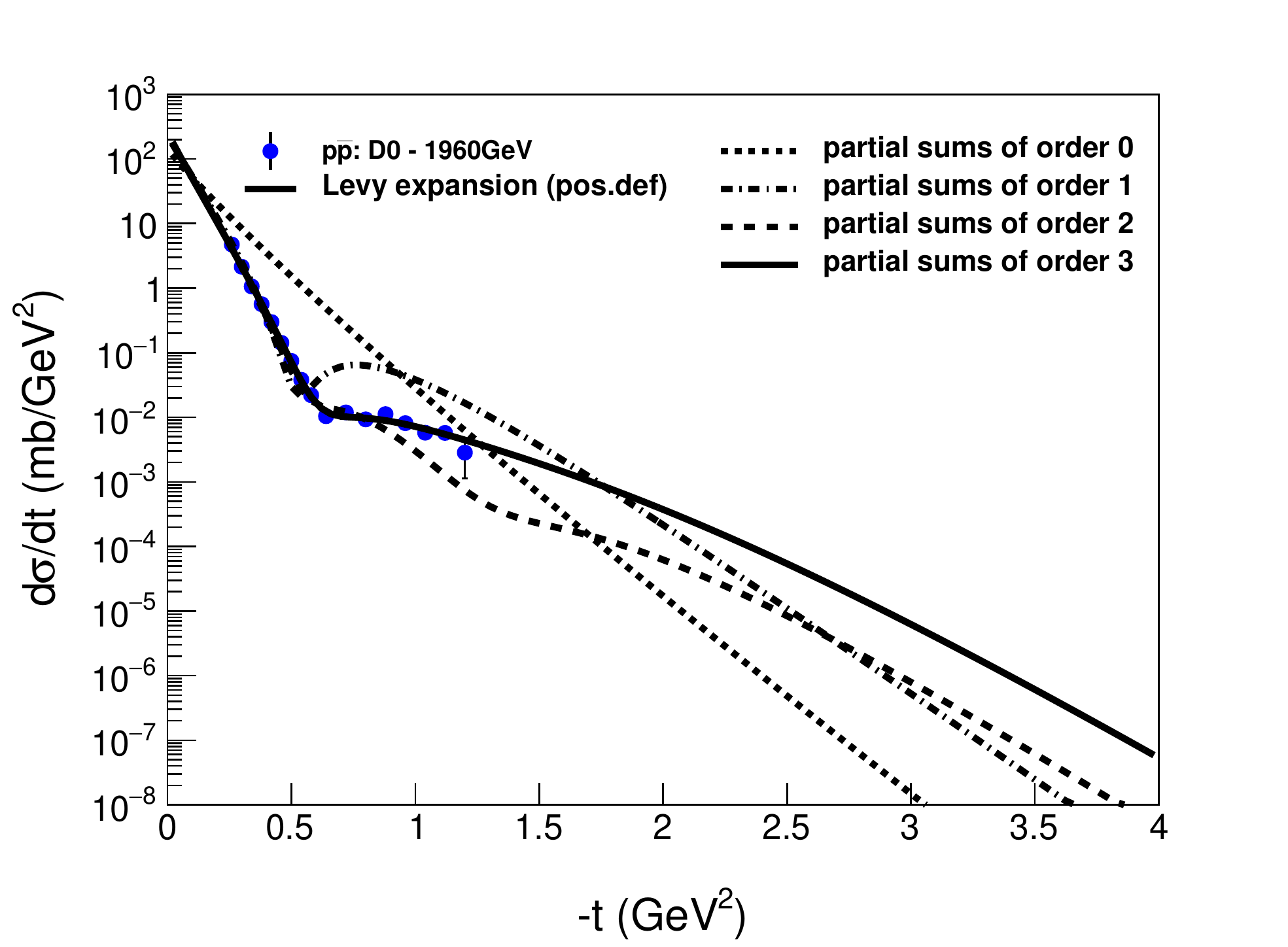}}
\end{minipage}
\caption{
Partial sums from leading order to fourth order L\'evy expansion, as compared to the TOTEM data  from Ref.~\cite{Antchev:2013gaa}
at $\sqrt{s} = 7 $ TeV $pp$ (left) and to D0 data from Ref.~\cite{Abazov:2012qb} on $p\overline{p}$ elastic scattering at $\sqrt{s} = 1.96$
TeV (right) elastic scattering.
}
\label{f:dsigdt-partial-sums}
\end{figure*}

\section{Observables}
\label{s:Observables}

Observables can be calculated from Eq.~(\ref{e:dsigmadt-Tel}). In what follows, we present the results for the fixed $b_0 = 0$.

The total cross-section is obtained, using the optical theorem, as follows
\begin{equation}
\sigma_{\rm tot} \equiv 2\,{\rm Im}\, T_{\rm el}(\Delta=0) \, =  2\,\sqrt{4\pi A}\,
	\left(1 + \sum_{i = 1}^\infty a_i l_i (0|\alpha) \right)\, .
    \label{e:sigmatot}
\end{equation}
The elastic cross-section, $\sigma_{\rm el}$ can be obtained using the orthonormality, 
Eq.~(\ref{e:lj}) as
\begin{equation}
     \sigma_{\rm el} \, = \, \int_{-t=0}^\infty dt \frac{d\sigma}{dt} \, = \, 
     \frac{A}{R^2}\left[\frac{1}{\alpha} \Gamma\left(\frac{1}{\alpha}\right) + \sum_{i= 1}^\infty (a_i^2 + b_i^2) \right]\, .
     \label{e:sigmael}
\end{equation}
The following $t$-dependent functions have also been analyzed:
\begin{itemize}
    \item The four-momentum transfer dependent elastic slope $B(t)$, defined as
\begin{equation}
	B(t) = \frac{d}{dt}\left(\ln \frac{d\sigma}{dt} \right)\,.
	\label{e:Bslope}
\end{equation}
    \item The four-momentum dependent $\rho(t)$, defined as the ratio of the real to  imaginary parts of $T_{\rm el}$
\begin{equation}
\rho(t)\equiv \frac{{\rm Re}\, T_{\rm el}(t)}{{\rm Im}\, T_{\rm el}(t)} =
	- \left. \frac{\sum_{i = 1}^\infty b_i l_i (z|\alpha)}{1+\sum_{i = 1}^\infty a_i l_i (z|\alpha)}\right|_{z= t R^2} .
    \label{e:rho}
\end{equation}
The value of $\rho(t)$ at $t=0$ can be measured in the Coulomb-nuclear 
interference region and in this work we refer to this value as $\rho_0=\rho(t=0)$.
    \item The first definition for the nuclear phase $\phi_1(t)$, that can be introduced as 
\begin{equation}
	T_{\rm el}(t) = 	|T_{\rm el}(t) | \exp\left[i \phi_1(t)\right] .
\end{equation}
An alternative definition was used recently by the TOTEM collaboration, corresponding to the
principal value of the nuclear phase, that reads as
\begin{equation}
 \phi_2(t) = \frac{\pi}{2} - \arctan \rho(t) .\label{e:phi-princip}
\end{equation}
If the nuclear phase $\phi_1(t)$ satisfies $0 \le \phi_1(t)\le \pi$, then the above two definitions are equivalent. However, for complex arguments, $\arctan(z)$ has branch cut discontinuities on the complex plane hence in general the two definitions
$\phi_1(t)$ and $\phi_2(t)$ are inequivalent, as detailed recently in Ref.~\cite{Csorgo:2018uyp}. In the present work we plot  the principal value of the nuclear phase, that corresponds to the definition with  Eq.~(\ref{e:phi-princip}), which 
by definition satisfies  $0 \le \phi_2(t)\le \pi$.
    \item We shall also study the convergence properties of  the shadow profile function, 
\begin{eqnarray}
	P(b) \equiv 1-\left|e^{-\Omega(b)}\right|^2 = [2-{\rm Im}\,t_{\rm el}(b)]{\rm Im}\,t_{\rm el}(b) - [{\rm Re}\,t_{\rm el}(b)]^2 \,,
    \label{e:shadow}
\end{eqnarray}
where $\Omega(b)$ is the complex-valued opacity function. These functions can be
represented in terms of the impact-parameter dependent elastic amplitude $t_{\rm el}(b)$, see for example Ref.~\cite{Csorgo:2018uyp}:
\begin{equation}
    t_{\rm el}(b) \, = \, \int \frac{d^2\Delta}{(2\pi)^2}\, e^{-i{\bm \Delta}{\bm b}}\,T_{\rm el}(\Delta)\, = \, i\left[ 1 - e^{-\Omega(b)} \right]\,, \quad \Delta\equiv|{\bm \Delta}|\,, \quad b\equiv|{\bm b}|\, .\label{e:telb}
\end{equation}

\end{itemize}

\section{Convergence properties}
\label{s:convergence}

The L\'evy series (\ref{e:dsigmadt-Tel}) enables one to precisely and model-independently characterise the $t$-dependence of the elastic cross-section $\frac{d\sigma}{dt}$ not only in the diffractive cone, but also significantly away from it, namely, in a vicinity of the diffractive minimum as well as beyond it in the large $t$ regime (diffractive tail). This is achieved by a fit of Eq.~(\ref{e:dsigmadt-Tel}) to the existing data controlling the fit quality and the convergence of the L\'evy ansatz to the data points in each particular region of 4-momentum transfers.

We adopted the following procedure: a fourth order L\'evy polynomial fit was performed to the measured $\frac{d\sigma}{dt}$ data for elastic $pp$ scattering.
The expansion coefficients $(a_i,b_i)$ are determined for $i=1,... ,i_{\rm max} = 4$. 
On the subsequent plots, the contributions are  summed only up to a given order $j \le i_{\rm max}$. Thus a procedure not unlike to a  Taylor series expansion is defined and the inclusion of each subsequent order enables us to see
how the series converges to the measurements on a more and more extended domain.
As described in Ref.~\cite{Csorgo:2018uyp},  good quality fits of elastic $pp$ scatterings were achieved for almost all of the published data sets, using $i_{ \rm max} = 4$. To fit the  $p\bar p$ datasets, that were measured in a more limited $t$-range and with reduced statistics, third order L\'evy expansions with $i_{ \rm max} = 3$ were sufficient.
\begin{figure*}[!h]
\begin{minipage}{0.495\textwidth}
 \centerline{\includegraphics[width=1.0\textwidth]{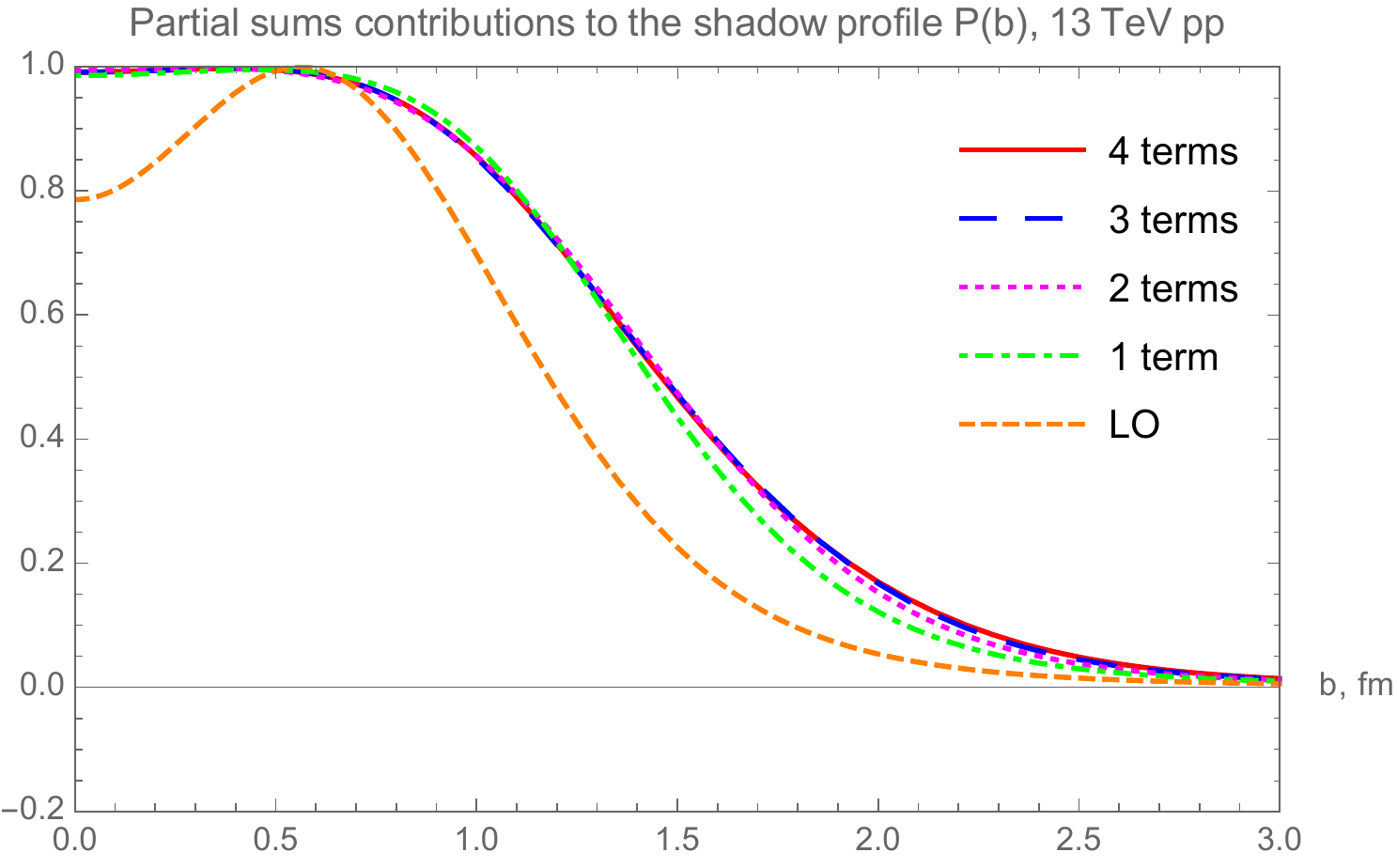}}
\end{minipage}
\begin{minipage}{0.495\textwidth}
 \centerline{\includegraphics[width=1.0\textwidth]{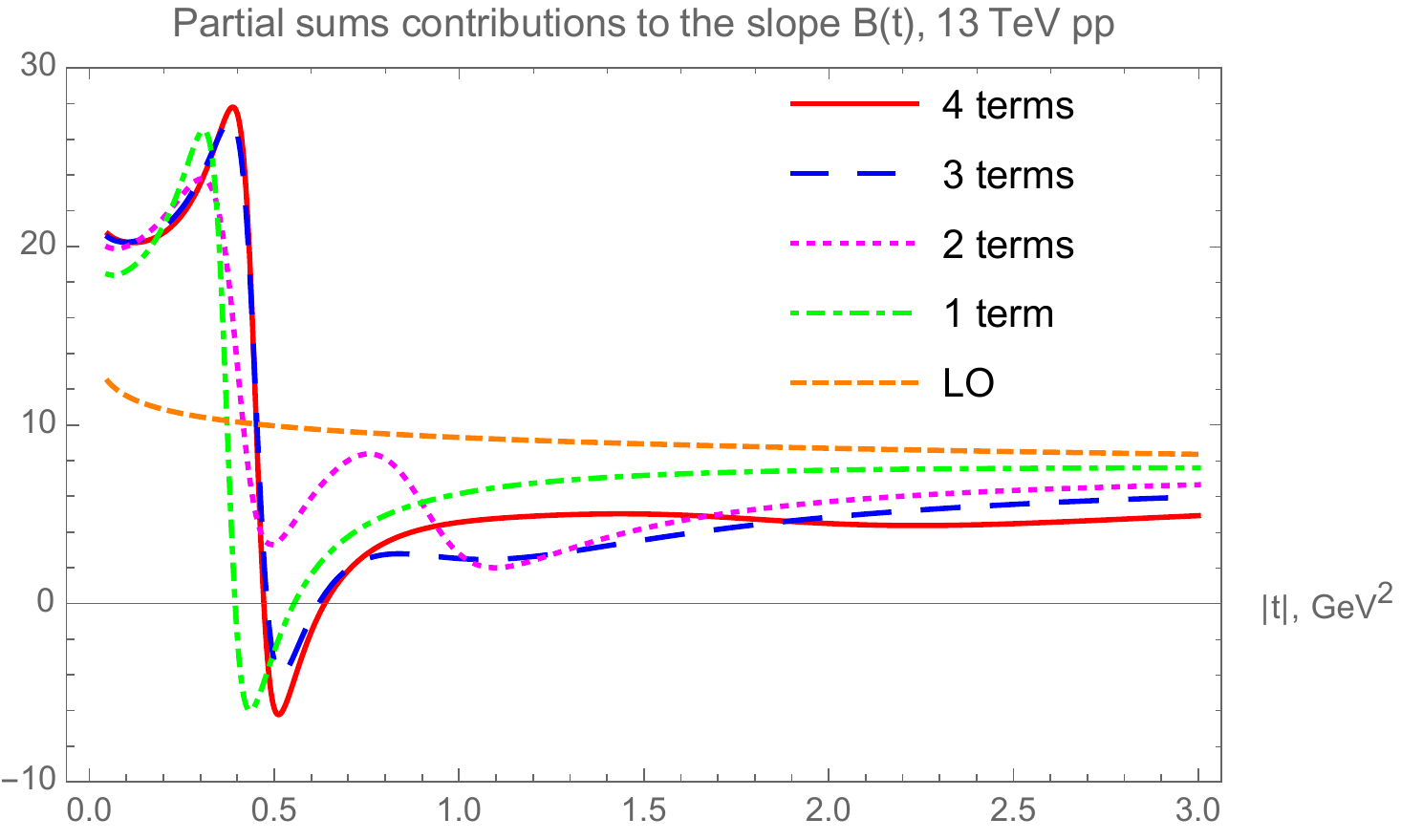}}
\end{minipage}
\caption{
The shadow profile (left) and the elastic slope (right) with partial sums from leading order 
to fourth order L\'evy expansion, fitted to the TOTEM data at $\sqrt{s} = 13 $ TeV $pp$ elastic scattering.
}
\label{f:Shadow-13-TeV-partial-sums}
\end{figure*}

As we demonstrate below, these fits converge rather fast to the data when increasing the order of the L\'evy expansion from zeroth to the fourth order. 

As an example of the convergence property, in Fig.~\ref{f:dsigdt-partial-sums}, we illustrate partial sums contributions to the L\'evy series (\ref{e:dsigmadt-Tel}) for the differential elastic cross-section against the TOTEM 7 TeV (left) and D0 1.96 TeV (right) data sets. 
As Figure~\ref{f:dsigdt-partial-sums} indicates, the zeroth order term fixes the total cross-section, i. e. the contribution at $t=0$
only, it apparently  insufficient even at small $t$ to approximate the shape of the distribution. The first order L\'evy expansions is required to describe $B(t=0)$ and the beginining of the diffractive cone. Although the first order contribution  exhibits a structure with a diffractive dip, the position and the size of the dip is not yet picked up correctly. 
The second order partial sum  is needed for getting the position of the diffractive minimum correctly, while the third order term fixes the position and the magnitude of the diffractive maximum.  The fourth order terms are necessary to describe the data also well beyond the diffractive minimum and maximum. 
\begin{figure*}[!h]
\begin{minipage}{0.495\textwidth}
 \centerline{\includegraphics[width=1.0\textwidth]{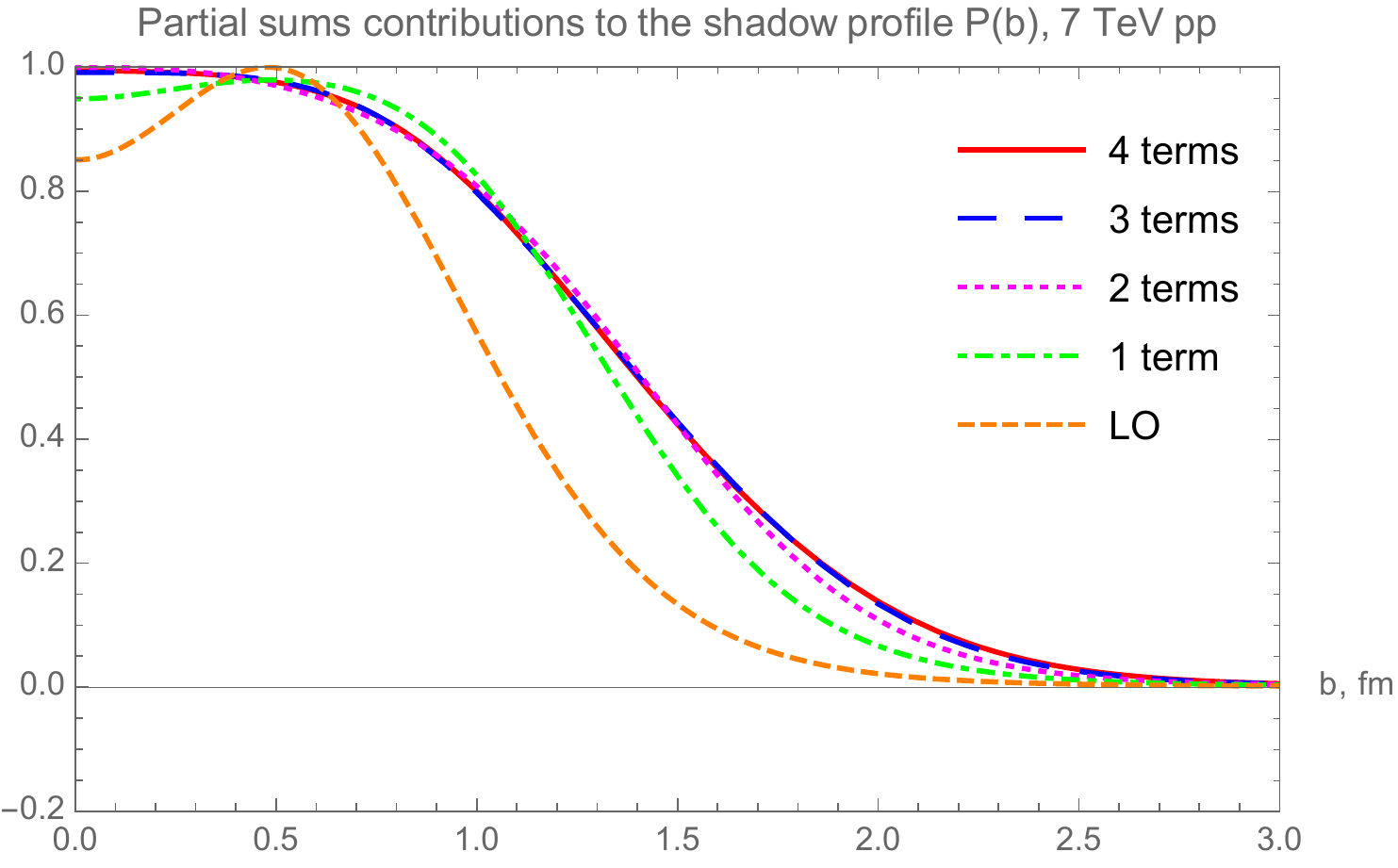}}
\end{minipage}
\begin{minipage}{0.495\textwidth}
 \centerline{\includegraphics[width=1.0\textwidth]{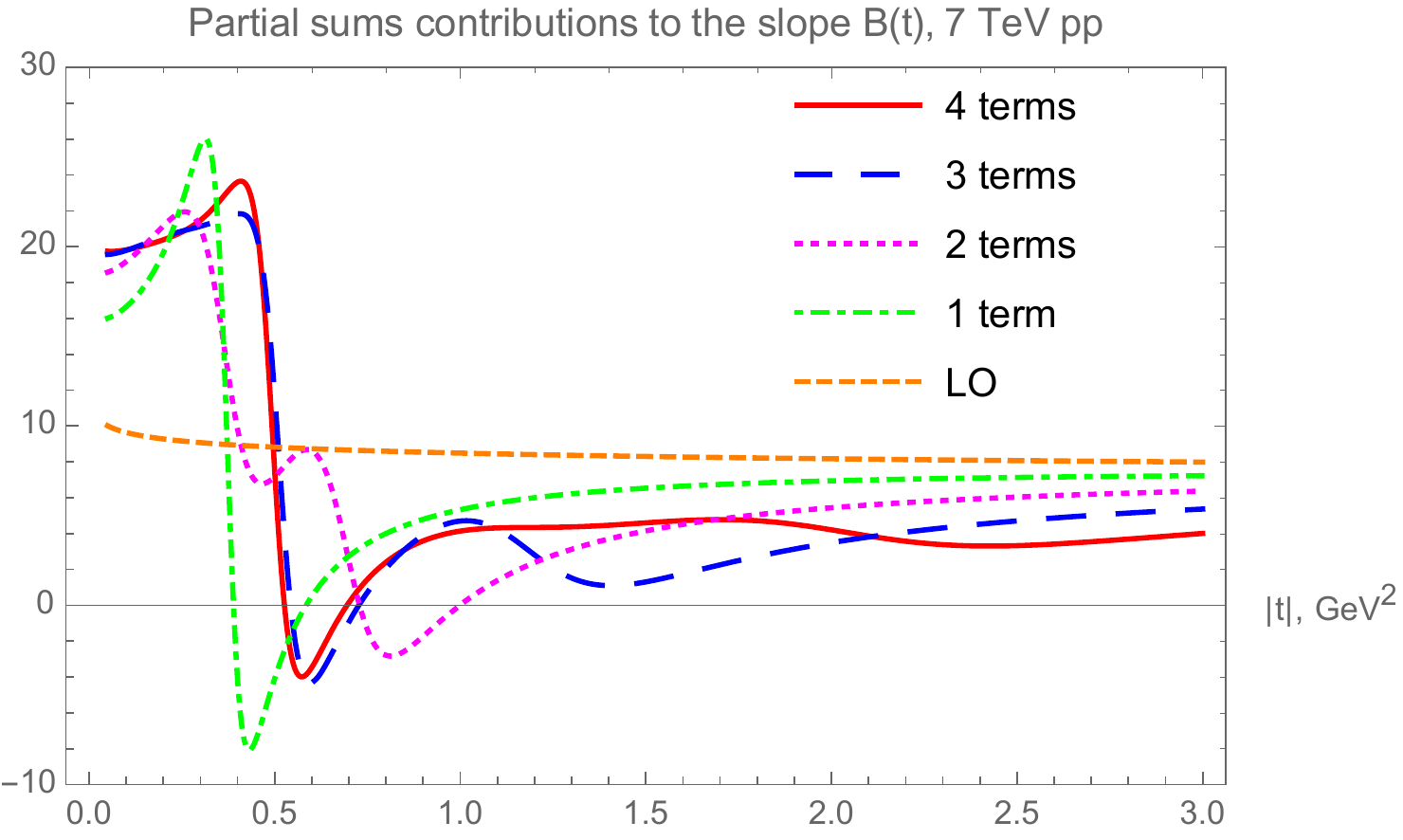}}
\end{minipage}
\caption{
The shadow profile (left) and the elastic slope (right) with partial sums from leading order 
to fourth order L\'evy expansion, fitted to the TOTEM data at $\sqrt{s} = 7 $ TeV $pp$ elastic scattering.
}
\label{f:Shadow-7-TeV-partial-sums}
\end{figure*}

While at zeroth order the real part of the elastic amplitude is found to be vanishing in our current approach, it may get (re)generated starting from the first order partial sums, and appears to be necessarily small. Remarkably, the considered quality fits
shown for example in Fig.~\ref{f:b0} for the 13 TeV elastic $pp$ scattering data of TOTEM enable us to reproduce the measured real-to-imaginary parts ratio $\rho(t)$ at $t=0$ with an excellent precision.

\begin{figure*}[!h]
\begin{minipage}{0.495\textwidth}
 \centerline{\includegraphics[width=1.0\textwidth]{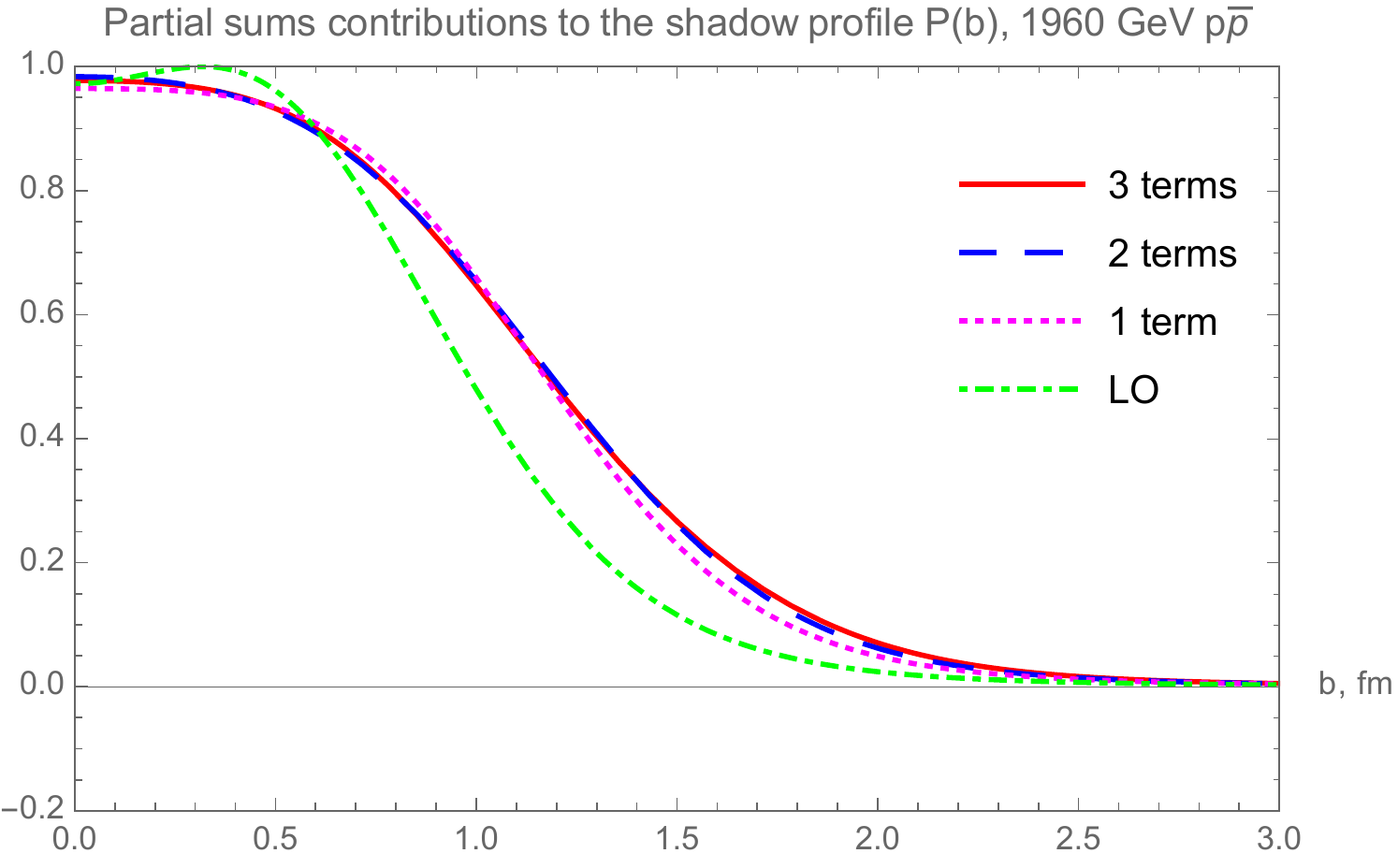}}
\end{minipage}
\begin{minipage}{0.495\textwidth}
 \centerline{\includegraphics[width=1.0\textwidth]{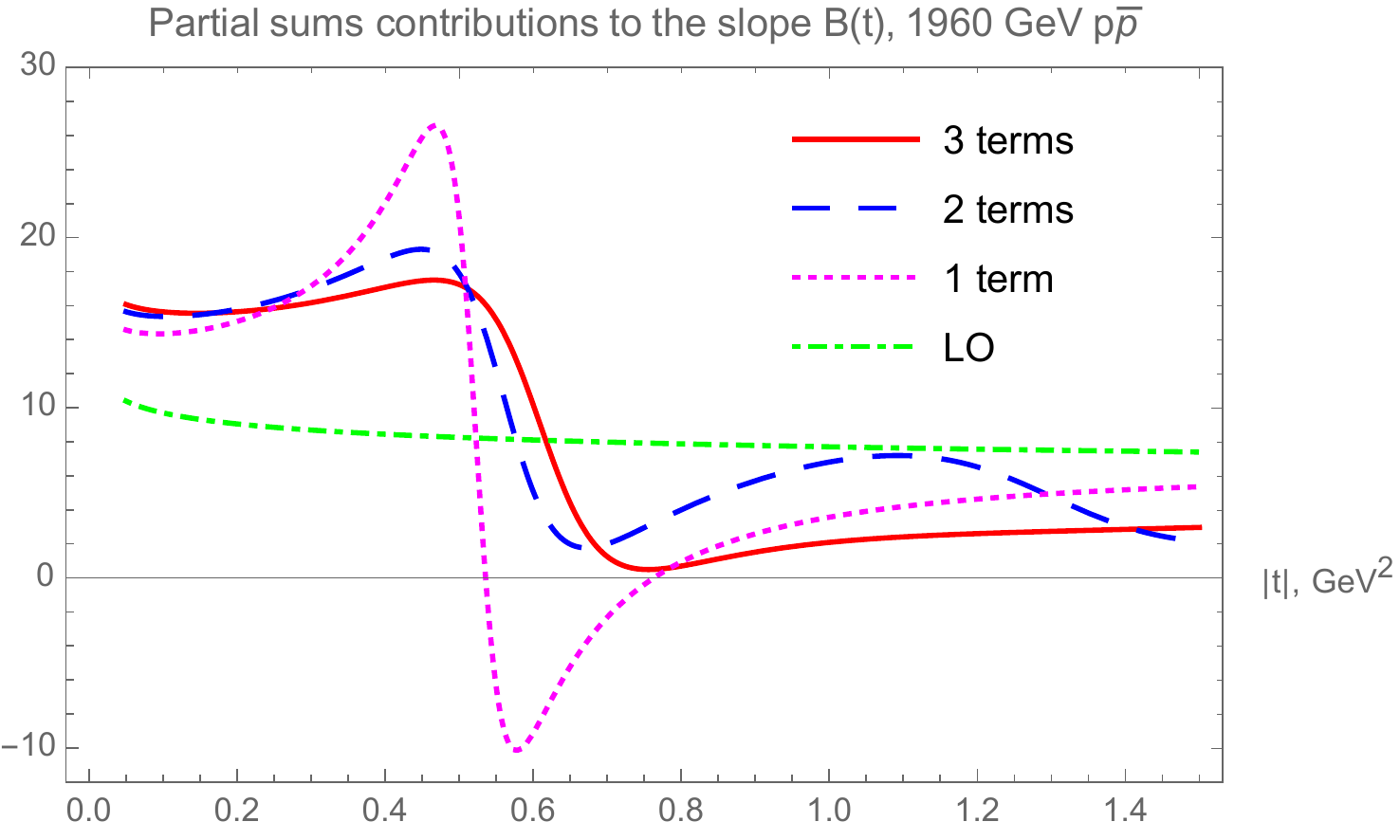}}
\end{minipage}
\caption{
The shadow profile (left) and the elastic slope (right) with partial sums from leading order 
to third order L\'evy expansion fitted to the D0 data at $\sqrt{s} = 1.96 $ TeV $p\bar p$ elastic scattering.
}
\label{f:Shadow-1_96-TeV-partial-sums}
\end{figure*}

In Figs.~\ref{f:Shadow-13-TeV-partial-sums}, \ref{f:Shadow-7-TeV-partial-sums} and \ref{f:Shadow-1_96-TeV-partial-sums} we show the convergence properties of the shadow profile (left panels) and the elastic slope (right panels) for L\'evy fits of the TOTEM data at 13 TeV, 7 TeV ($pp$ collisions) and D0 data at 1.96 TeV ($p\bar p$ collisions) at different consecutive orders as previously illustrated in Fig.~\ref{f:dsigdt-partial-sums}, respectively. 

The convergence of partial sums contributions to the shadow profile turns out to be faster than that for the elastic slope. Namely, already starting from the second L\'evy order the result for $P(b)$ remains fairly stable at large $b$ for all the considered data sets, but higher order terms are needed for a precise result at small values of $b$. For $pp$ collisions TOTEM data at 13 and 7 TeV, the behavior of the elastic slope stabilizes in a vicinity of the diffractive dip/bump structure only at third L\'evy order, in consistency with the observations made from Fig.~\ref{f:dsigdt-partial-sums}. So, for all TOTEM data sets the minimal preferred order enabling to extend the data description significantly beyond the dip/bump structure is the fourth. Due to a lack of data in the large-$t$ region for $p\bar p$ collisions at 1.96 TeV, we limit ourselves to the third order that provides the results for $B(t)$ sufficiently stable in the ``shoulder'' region.
\begin{figure*}[!h]
\begin{minipage}{0.495\textwidth}
 \centerline{\includegraphics[width=1.0\textwidth]{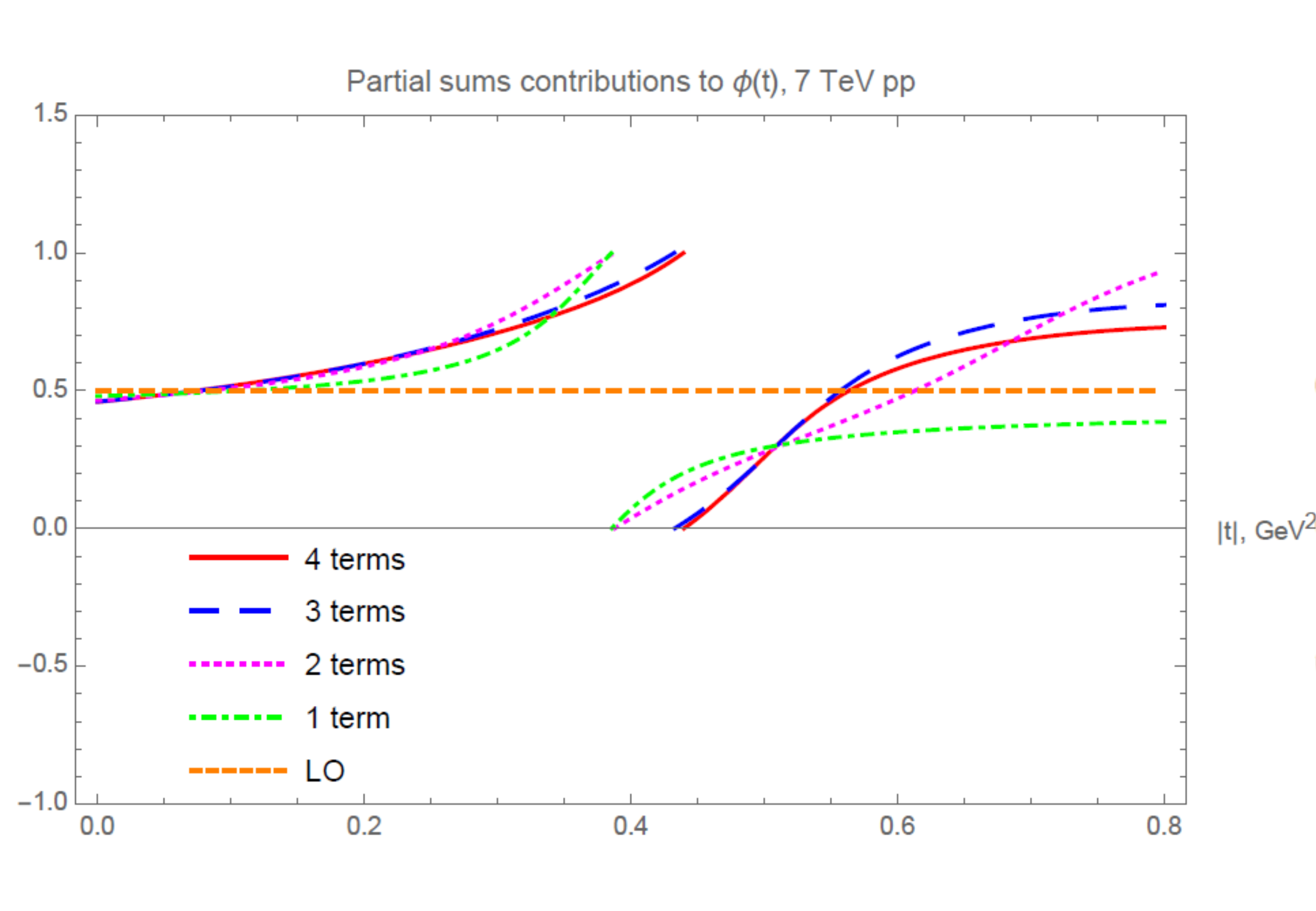}}
\end{minipage}
\begin{minipage}{0.495\textwidth}
 \centerline{\includegraphics[width=1.0\textwidth]{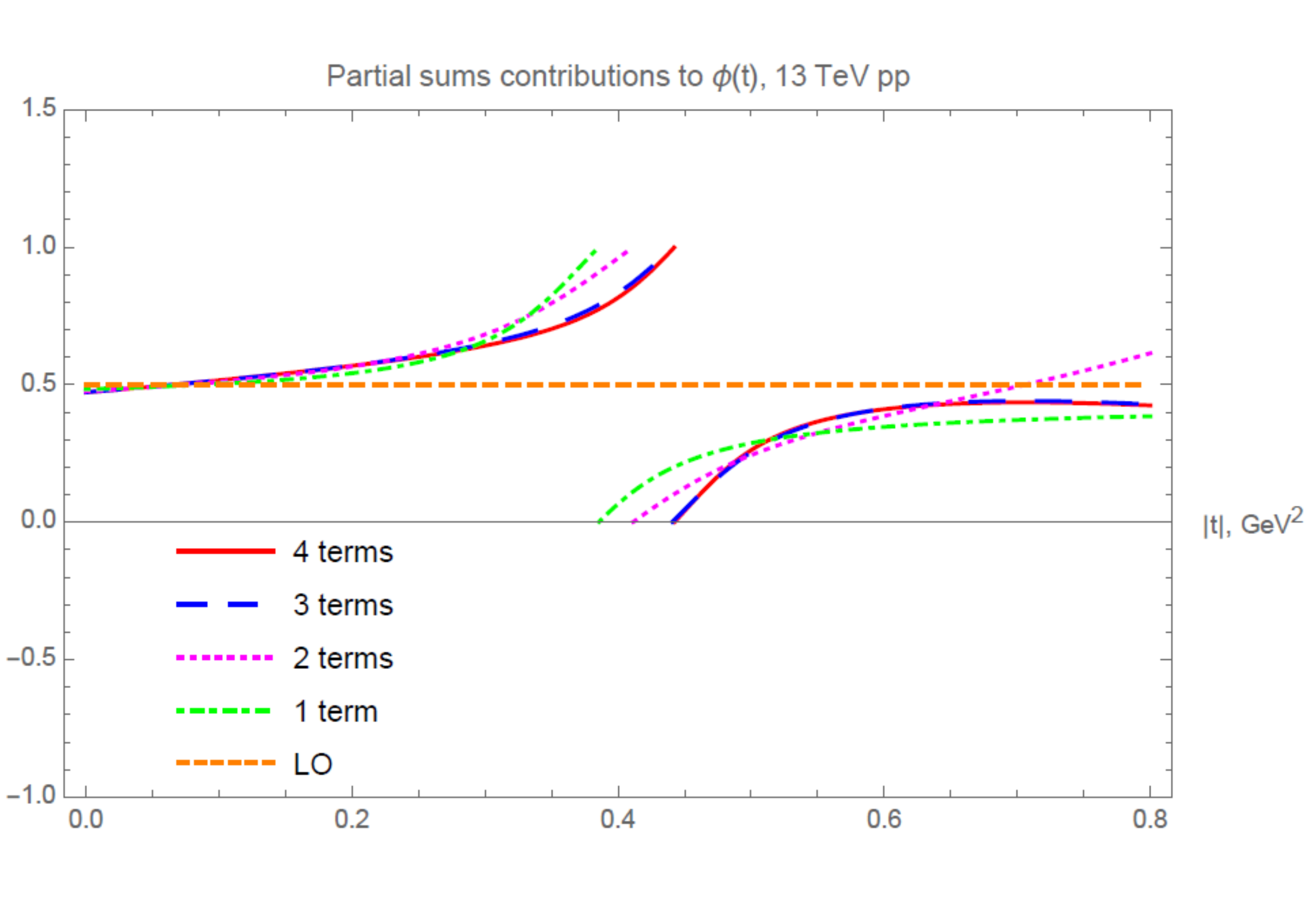}}
\end{minipage}
\caption{
The principal value of the nuclear phase $\phi_2(t)$ (in units of $\pi$) with the contributions of the partial sums from the leading order 
to fourth order L\'evy expansion, as fitted to the TOTEM $pp$ elastic scattering data 
at $\sqrt{s} = 7 $ TeV (left panel) and at $\sqrt{s} = 13 $ TeV (right panel), respectively. Note the stability of the point $\phi_2(t) = \pi$ for increasing order of the expansion, and for changing the energy of the collision.
}
\label{f:Phase-13-TeV-partial-sums}
\end{figure*}

Finally, in Fig.~\ref{f:Phase-13-TeV-partial-sums} we demonstrate the corresponding
convergence property of $\phi_2(t)$, the principal value of the nuclear phase for $\sqrt{s} = 13 $ TeV $pp$ elastic scattering. We observe that there is no significant difference between the third- and the fourth-order L\'evy expansion results indicating a good stability property 
of this function in the range of the diffractive minimum, where $\phi_2(t_{\rm dip}) = \pi$.

\section{Odderon}

The concept of Odderon exchange corresponds to a crossing-odd term in the $pp$ elastic scattering amplitude. This concept was introduced by Lukaszuk and Nicolescu in 1973~\cite{Lukaszuk:1973nt}. 

The recent TOTEM  results~\cite{Antchev:2017dia,Antchev:2017yns}
generated a burst of high level  and intense theoretical debate about the 
correct interpretation of these data, see Refs.~\cite{Samokhin:2017kde,Khoze:2017swe,Petrov:2018xma,Khoze:2018bus,Goncalves:2018yxc,Shabelski:2018jfq,Broilo:2018els,Lebiedowicz:2018eui,Martynov:2018nyb,Troshin:2018ihb,Dremin:2018uwt,Broniowski:2018xbg,Khoze:2018kna}.
All possible extreme views were present among these first interpretative papers, 
including claims for a maximal Odderon effect~\cite{Martynov:2018nyb}  and claims of lack of any significant Odderon effects, Refs.~\cite{Shabelski:2018jfq,Khoze:2018kna}. A review of recent theoretical developments on possible Odderon effects is given in Refs.~\cite{Martynov:2018pye,Martynov:2018yas}.

At sufficiently high energies, the relative contribution from secondary Regge 
trajectories is suppressed, as they decay as negative powers of the colliding energy $\sqrt{s}$. The vanishing nature of these Reggeon contributions offers a direct way of extracting the Odderon as well as the Pomeron contributions, $T_{el}^{O}(s,t)$ and $T_{el}^{P}(s,t)$, respectively, from elastic scattering data at sufficiently high colliding energies.  Thus, the Odderon effects are expected to be detected clearly with measurements in the TeV energy range~\cite{Avila:2006wy,Martynov:2017zjz}.

In Ref.~\cite{Ster:2015esa}, the authors also argued that the LHC energy scale is already sufficiently large to suppress the Reggeon contributions, and they presented the $(s,t)$ dependent contributions of Odderon exchange to the differential and total cross-sections
at LHC energies. That analysis relied on a model-dependent, phenomenological extension of the 
Phillips-Barger model~\cite{Phillips:1974vt}, that focused on fitting the dip region
of elastic proton-proton scattering, but did not analyze in detail the tails and the cone region. The fitted model parameters of proton-proton and proton-antiproton reactions were extrapolated to exactly the same energies, and the results were recently confirmed
and extended in Ref.~\cite{Goncalves:2018nsp}. Similarly, Ref.~\cite{Lebiedowicz:2018eui} also argued that the currently highest LHC energy of $\sqrt{s} = $ 13 TeV is sufficiently high to see various Odderon contributions. In particular, the Pomeron and the Odderon contributions
can be extracted from the forward scattering amplitudes at sufficiently high energies as discussed, for example, in Refs.~\cite{Ster:2015esa,Csorgo:2018uyp}.
Elastic proton-proton and proton-antiproton scattering data were not measured at the same energy
in the TeV region so far.  However, we have identified two robust-looking features of the already performed
measurements, that  provided not only an Odderon signal, but they
also clearly indicated  the existence of two different sizes for some sub-structures inside the
protons, as imaged by elastic proton-proton scattering.

In particular, we found that clear, but indirect signals of Odderon effects are
present in the difference between the $t$-dependent nuclear slopes of elastic proton-proton and proton-antiproton scattering~\cite{Csorgo:2018uyp,Csorgo:2019rsr}. Our results for the existence
of a well-defined and negative minimum of the $B(t)$ functions for $pp$ reactions and a lack of significantly negative values of $B(t)$  in $p\overline{p}$ reactions at the TeV region has recently been  confirmed using the maximal Odderon model of Martynov and Nicolescu in Ref.~\cite{Martynov:2018sga}. From the experimental point of view,
$B(t)$ is straightforward to measure and can be used for an experimental search for  Odderon effects 
independently of L\'evy expansion and imaging results~\cite{Csorgo:2019fbf}. 

In addition, we  have also identified a clear difference  between the principal values of the nuclear phase, $\phi_2(t)$ of proton-proton and proton-antiproton collisions in the TeV energy range~\cite{Csorgo:2018uyp,Csorgo:2019rsr}. 
However, the $t$-dependence of the nuclear phase is rather difficult, close to impossible to access experimentally, in particular independently of the L\'evy imaging methods.

Both of these Odderon effects are  obtained with a convergent series expansion, and are stable for higher order L\'evy expansion coefficients, as indicated on the right panels of 
Figs.~\ref{f:Shadow-13-TeV-partial-sums}, \ref{f:Shadow-7-TeV-partial-sums} and \ref{f:Shadow-1_96-TeV-partial-sums} in case of the nuclear slope $B(t)$ and on Fig.~\ref{f:Phase-13-TeV-partial-sums} in case of the principal value of the nuclear phase $\phi_2(t)$. 

\section{Summary and conclusions}

Our analysis in Refs.~\cite{Csorgo:2018uyp,Csorgo:2018ruk,Csorgo:2019rsr} has been primarily motivated by the search for Odderon effects. We have identified two independent Odderon effects in TOTEM differential cross-section measurements. The comparision of $B(t)$ for $pp$ and $p\overline{p}$ reactions at exactly the same $\sqrt{s}$ in the TeV region is one of the most promising channel for the experimental observation of Odderon effects.

One of the most obvious but nevertheless striking feature  of the elastic $pp$ scattering at TeV energies
is that  the differential cross-section has a unique, single minimum. In  multiple diffractive scattering theory,
single diffractive minimum may  appear in symmetric collisions of composite objects if and only if the colliding systems have  two internal substructures~\cite{Czyz:1969jg}. This suggests that the quark-diquark picture
of elastic proton-proton collisions, where a diquark that acts as a single unit in elastic scattering even at the LHC energies, formulated in terms of the real extended Bialas-Bzdak model of elastic proton-proton scattering~\cite{Bialas:2006qf,Nemes:2015iia}, may indeed capture correctly some of the most fundamental properties of elastic proton-proton collisions at the LHC energies.

One of our most surprising result was a clear-cut evidence for two different sub-structures inside the protons, as detailed in Ref.~\cite{Csorgo:2018uyp}.  We determined the significance of these substructure effects and estimated the sizes of these sub-structures and their contributions to the total and elastic  proton-proton cross-section in Ref.~\cite{Csorgo:2018ruk}.

\section*{Acknowledgments}

We would like to thank S. Giani,  G. Gustafson, V. Khoze, F. Nemes, B. Nicolescu and K. \"Osterberg for inspiring and clarifying discussions.
T. Cs. expresses his gratitude to the Organizers of ISMD 2018 for the opportunity to present these results, for partial support and for an outstanding, inspiring and useful meeting. T. Cs. was partially supported by the Hungarian NKIFH grants FK-123842 and FK-123959 
and the EFOP-3.6.1-16-2016-00001 grants (Hungary). R.P.~was partially supported by the Swedish Research Council grants, contract numbers 
621-2013-4287 and 2016-05996, CONICYT grant MEC80170112, and by the Ministry of Education, Youth and Sports of the Czech Republic project LT17018. 
This research was partially supported by the THOR project, COST Action CA15213 of the European Union.


\begin{thebibliography}{99}

\bibitem{Csorgo:2018uyp} 
  T.~Cs\"org\H{o}, R.~Pasechnik and A.~Ster,
  Eur.\ Phys.\ J.\ C {\bf 79}, no. 1, 62 (2019)

\bibitem{Csorgo:2018ruk} 
  T.~Cs\"org\H{o}, R.~Pasechnik and A.~Ster,
  arXiv:1811.08913 [hep-ph]

\bibitem{Csorgo:2019rsr} 
  T.~Cs\"org\H{o}, R.~Pasechnik and A.~Ster,
  arXiv:1902.00109 [hep-ph]

\bibitem{Antchev:2017dia} 
  G.~Antchev {\it et al.} [TOTEM Collaboration],
  Eur.\ Phys.\ J.\ C {\bf 79}, no. 2, 103 (2019)

\bibitem{Antchev:2017yns} 
  G.~Antchev {\it et al.} [TOTEM Collaboration],
  arXiv:1812.04732 [hep-ex]
  
\bibitem{Antchev:2018edk} 
  G.~Antchev {\it et al.} [TOTEM Collaboration],
  arXiv:1812.08283 [hep-ex]
  
\bibitem{Antchev:2018rec} 
  G.~Antchev {\it et al.} [TOTEM Collaboration],
  arXiv:1812.08610 [hep-ex]

\bibitem{Abazov:2012qb} 
  V.~M.~Abazov {\it et al.} [D0 Collaboration],
  Phys.\ Rev.\ D {\bf 86}, 012009 (2012)
  
\bibitem{Antchev:2015zza} 
  G.~Antchev {\it et al.} [TOTEM Collaboration],
  Nucl.\ Phys.\ B {\bf 899}, 527 (2015)
  
\bibitem{DeKock:2012gp} 
  M.~B.~De Kock, H.~C.~Eggers and T.~Cs\"org\H{o},
  PoS WPCF {\bf 2011}, 033 (2011)
  
\bibitem{Novak:2016cyc} 
  T.~Nov\'ak, T.~Cs\"org\H{o}, H.~C.~Eggers \& M.~De Kock,
  Acta Phys. Pol. Supp. {\bf 9}, 289 (2016)
  
\bibitem{Csorgo:1999wx} 
  T.~Cs\"org\H{o} and S.~Hegyi,
  Phys.\ Lett.\ B {\bf 489}, 15 (2000)

\bibitem{Antchev:2013gaa} 
  G.~Antchev {\it et al.} [TOTEM Collaboration],
  EPL {\bf 101}, no. 2, 21002 (2013)

\bibitem{Lukaszuk:1973nt} 
  L.~Lukaszuk and B.~Nicolescu,
  Lett.\ Nuovo Cim.\  {\bf 8}, 405 (1973)

\bibitem{Samokhin:2017kde} 
  A.~P.~Samokhin and V.~A.~Petrov,
  Nucl.\ Phys.\ A {\bf 974}, 45 (2018)

\bibitem{Khoze:2017swe} 
  V.~A.~Khoze, A.~D.~Martin and M.~G.~Ryskin,
  Phys.\ Rev.\ D {\bf 97}, no. 3, 034019 (2018)
  
\bibitem{Petrov:2018xma} 
  V.~A.~Petrov,
  Eur.\ Phys.\ J.\ C {\bf 78}, no. 3, 221 (2018)
  Erratum: [Eur.\ Phys.\ J.\ C {\bf 78}, no. 5, 414 (2018)]

\bibitem{Khoze:2018bus} 
  V.~A.~Khoze, A.~D.~Martin and M.~G.~Ryskin,
  Phys.\ Lett.\ B {\bf 780}, 352 (2018)
  
\bibitem{Goncalves:2018yxc} 
  V.~P.~Gonçalves and B.~D.~Moreira,
  Phys.\ Rev.\ D {\bf 97}, no. 9, 094009 (2018)

\bibitem{Shabelski:2018jfq} 
  Y.~M.~Shabelski and A.~G.~Shuvaev,
  Eur.\ Phys.\ J.\ C {\bf 78}, no. 6, 497 (2018)

\bibitem{Broilo:2018els} 
  M.~Broilo, E.~G.~S.~Luna and M.~J.~Menon,
  Phys.\ Lett.\ B {\bf 781}, 616 (2018)
  
\bibitem{Lebiedowicz:2018eui} 
  P.~Lebiedowicz, O.~Nachtmann and A.~Szczurek,
  Phys.\ Rev.\ D {\bf 98}, 014001 (2018)

\bibitem{Martynov:2018nyb} 
  E.~Martynov and B.~Nicolescu,
  Phys.\ Lett.\ B {\bf 786}, 207 (2018)
  
\bibitem{Troshin:2018ihb} 
  S.~M.~Troshin and N.~E.~Tyurin,
  Mod.\ Phys.\ Lett.\ A {\bf 33}, no. 35, 1850206 (2018)
  
\bibitem{Dremin:2018uwt} 
  I.~M.~Dremin,
  Universe {\bf 4}, no. 5, 65 (2018)

\bibitem{Broniowski:2018xbg} 
  W.~Broniowski, L.~Jenkovszky, E.~Ruiz Arriola and I.~Szanyi,
  Phys.\ Rev.\ D {\bf 98}, (7), 074012 (2018)
  

\bibitem{Khoze:2018kna} 
  V.~A.~Khoze, A.~D.~Martin and M.~G.~Ryskin,
  Phys.\ Lett.\ B {\bf 784}, 192 (2018)

\bibitem{Martynov:2018pye} 
  E.~Martynov and B.~Nicolescu,
  arXiv:1811.07635 [hep-ph]
  
\bibitem{Martynov:2018yas} 
  E.~Martynov and B.~Nicolescu,
  arXiv:1810.08930 [hep-ph]

\bibitem{Avila:2006wy} 
  R.~Avila, P.~Gauron and B.~Nicolescu,
  Eur.\ Phys.\ J.\ C {\bf 49}, 581 (2007)

\bibitem{Martynov:2017zjz} 
  E.~Martynov and B.~Nicolescu,
  Phys.\ Lett.\ B {\bf 778}, 414 (2018)

\bibitem{Ster:2015esa} 
  A.~Ster, L.~Jenkovszky and T.~Cs\"org\H{o},
  Phys.\ Rev.\ D {\bf 91}, no. 7, 074018 (2015)
  
\bibitem{Phillips:1974vt} 
  R.~J.~N.~Phillips and V.~D.~Barger,
  Phys.\ Lett.\  {\bf 46B}, 412 (1973)
  
\bibitem{Goncalves:2018nsp} 
  V.~P.~Gonçalves and P.~V.~R.~G.~Silva,
  arXiv:1811.12250 [hep-ph]


\bibitem{Martynov:2018sga} 
  E.~Martynov and B.~Nicolescu,
  arXiv:1808.08580 [hep-ph]
  
\bibitem{Csorgo:2019fbf} 
 T. Cs\"org\H{o}, for the TOTEM Collaboration,
  arXiv:1903.06992 [hep-ex] \\
Proc. XLVIII ISMD, Singapore, September 3-7, 2018

\bibitem{Czyz:1969jg} 
  W.~Czyz and L.~C.~Maximon,
  Annals Phys.\  {\bf 52}, 59 (1969)

\bibitem{Bialas:2006qf} 
  A.~Bialas and A.~Bzdak,
  Acta Phys.\ Polon.\ B {\bf 38}, 159 (2007)

\bibitem{Nemes:2015iia} 
  F.~Nemes, T.~Cs\"{o}rg\H{o} and M.~Csan\'ad,
  Int.\ J.\ Mod.\ Phys.\ A {\bf 30}, no. 14, 1550076 (2015)

\end{thebibliography}
\end{document}